\documentstyle[aps,preprint]{revtex}
\begin{document}

\large

\vspace*{.1in}

\begin{center}
\rm\Large
Position-momentum local realism violation of the Hardy type
\end{center}

\vspace*{1ex}

\begin{center}
{Bernard Yurke$^1$, Mark Hillery$^2$, and David Stoler$^1$}
\end{center}

\begin{center}
{\large \em $^1$Bell Laboratories, Lucent Technologies, Murray Hill, NJ 07974}
\end{center}

\begin{center}
{\large \em $^2$ Hunter College of the City University of New York,
695 Park Ave. New York, NY 10021}
\end{center}

\begin{center}
\today
\end{center}

\begin{flushleft}
We show that it is, in principle, possible to perform
local realism violating experiments of the Hardy type in which only
position and momentum measurements are made on two particles
emanating from a common source.  In the optical domain, homodyne 
detection of the in-phase and out-of-phase amplitude components
of an electromagnetic field is analogous to position and momentum
measurement.  Hence, local realism violations of the Hardy type
are possible in optical systems employing
only homodyne detection.
\end{flushleft}

\vspace*{2ex}


\vspace*{1ex}

file:{\small stoler/Local-realism/xpprl971106.tex}

\newpage

As an example to support their contention that quantum mechanics is
incomplete Einstein, Podolsky, and Rosen (EPR) \cite{einstein35} exhibited
a quantum mechanical wave function, describing two particles emitted
from a common source, in which the positions and the momenta of the two particles
were strongly correlated.  This wave function described the situation in which
the measurement of the position of one of the particles would allow
one to predict with complete certainty the position of the other 
particle and the measurement of the momentum of one of the
particles would allow one to predict with complete certainty the momentum
of the other particle.  Because of these strong correlations even when
the particles were well-separated it was argued 
that each of the particles must have a definite position and a definite momentum 
even though a quantum mechanical wave function does not simultaneously ascribe
a definite position and a definite momentum to a particle.  Therefore, it was argued
that quantum mechanics is incomplete.  It was hoped that in the future a complete
theory could be devised in which a definite position and definite momentum
would be ascribed 
to each particle.  In 1992 the EPR Gedanken experiment was actually carried
out \cite{ou92} as a quantum optics experiment in which electromagnetic field 
analogues of position and momentum were measured on correlated photon states 
generated by parametric down-conversion \cite{reid88,reid89}.  The analogues 
of the position and momentum were the two quadrature amplitudes of the 
electromagnetic field measured via homodyne detectors 
\cite{yuen80,schumaker84,yurke87}.  A quantum mechanical state having
the properties of the state employed by EPR had been realized.  

However, since the work of Bell \cite{bell65} it has been known
that a complete theory, of the type EPR hoped for, capable of making the same 
predictions as quantum mechanics, does not exist \cite{clauser78}.  
A variety of experiments,
referred to as local realism violating experiments, have been proposed and 
performed, demonstrating that quantum mechanics is inherently at odds with 
classical notions about how effects propagate.  Most striking among the
proposals are the ``one event'' local realism violating experiments
devised by Greenberger Horne and Zeilinger (GHZ) \cite{greenberger89,mermin90,yurke92} 
and by Hardy \cite{hardy92,clifton92,yurke93}.  The Bell, GHZ, and Hardy experiments
that have been proposed  generally measure spin components or count particles, 
i.e., they employ observables that have a discrete spectrum.  There are, however, 
some examples in which continuous observables or a mixture of discrete and continuous
observables have been employed \cite{bell87,cetto85,yurke97,gilchrist98}.  
In fact, Bell
\cite{bell87} showed that position and momentum measurements on a pair of
particles in a state for which the Wigner function has negative regions
can give rise to local realism violating effects of the Clauser, Holt,
Horne, and Shimony type \cite{clauser69}.  Here we show that 
local realism violating effects of the Hardy type can be obtained through 
position and momentum measurements on a pair of particles prepared in the 
appropriate state.  Given that homodyne detection 
measurements of the two quadrature amplitude components of an electromagnetic 
field provide an optical analogue to position and momentum measurements, an optical 
experiment exhibiting local realism violations of the Hardy type can
be devised, provided the appropriate entangled state can be generated.

A local realism constraint on the positions and momenta measured
for a pair of particles emitted from a common source can be 
arrived at by regarding the detectors as responding to messages 
emitted by the source \cite{mermin90,yurke97}.  
The source does not know, ahead of time,
whether a position or a momentum measurement will be performed by
a given detector.  Hence, the instruction set emitted by the source
must tell the detectors what to do in either case.  The instruction
sets are conveniently labeled via the array $(\alpha_{x1}, \alpha_{x2};
\alpha_{p1}, \alpha_{p2})$ where $\alpha_{xi}$ and $\alpha_{pi}$ are
members of the set $\{+,-\}$. Here $\alpha_{xi}$ denotes whether detector 
$i$, measuring the position ${x_i}$ of particle $i$, will report the position
to be positive ($\alpha_{xi} = +$) or negative ($\alpha_{xi} = -$).
Similarly, $\alpha_{pi}$ denotes whether detector 
$i$, measuring the momentum ${p_i}$ of particle $i$, will report the momentum
to be positive ($\alpha_{pi} = +$) or negative ($\alpha_{pi} = -$).
The probability that a message of the form  $(\alpha_{x1}, \alpha_{x2};
\alpha_{p1}, \alpha_{p2})$ will be denoted as  $P(\alpha_{x1}, \alpha_{x2};
\alpha_{p1}, \alpha_{p2})$.  Let $P_{\beta_1 \beta_2}(\alpha_{\beta_1 1},
\alpha_{\beta_2 2})$, where $\beta_i \in \{x,p\}$, is the probability that
detector $1$ measuring $\beta_1$ reports $\alpha_{\beta_1 1}$ while
detector $2$ measuring $\beta_2$ reports $\alpha_{\beta_2 2}$.  For
example, $P_{xp}(+,-)$ denotes the joint probability that
detector 1 measuring position will report a positive position while
detector 2 measuring momentum will report a negative momentum.
In terms of the message probabilities, $P_{pp}(-,-)$ is
given by
\begin{eqnarray}
P_{pp}(-,-) & = & P(+,+;-,-) + P(+,-;-,-) \nonumber \\
& & + \ P(-,+;-,-) + P(-,-;-,-) \ .
\label{eq:2.1}
\end{eqnarray}
The joint probabilities provide the following bounds on the message
probabilities
\begin{eqnarray}
P(+,+;-,-) & \leq & P_{xx}(+,+) \label{eq:2.2} \ ,\\
P(+,-;-,-) & \leq & P_{px}(-,-) \label{eq:2.3} \ , \\
P(-,+;-,-) & \leq & P_{xp}(-,-) \label{eq:2.4} \ , 
\end{eqnarray}
and
\begin{equation}
P(-,-;-,-) \leq \min\{P_{xp}(-,-),P_{px}(-,-)\} \ . 
\label{eq:2.5}
\end{equation}
Applying these inequalities to Eq.~(\ref{eq:2.1}) yields the following
local realism constraint on the joint probabilities:
\begin{eqnarray}
P_{pp}(-,-) & \leq & P_{xx}(+,+) + P_{px}(-,-) + P_{xp}(-,-) \nonumber \\
& & + \ \min\{P_{xp}(-,-),P_{px}(-,-)\} \ . \label{eq:2.6}
\end{eqnarray}


If it is rigorously known that the probabilities on the right-hand side of 
the inequality (\ref{eq:2.6}) are all zero,
\begin{equation}
P_{xx}(+,+) = P_{xp}(-,-) = P_{px}(-,-) = 0 \ , 
\label{eq:3.1}
\end{equation}
then it follows, according to local realism, that $P_{pp}(-,-)$ is 
rigorously zero.  Thus, the appearance of a single event in which both
particles have negative momentum would violate local realism.
This situation, referred to as ``one event'' local realism violation,
of course, cannot be achieved in practice because with a finite
amount of data or the presence of spurious events it is impossible
to rigorously demonstrate, experimentally, that Eq.~(\ref{eq:3.1}) 
holds for a given physical system.  Nevertheless, if
the spurious event rate is sufficiently small, it is possible to 
demonstrate to a high degree of certainty with a finite amount of 
data that the inequality
Eq.~(\ref{eq:2.6}) is violated.

It is shown here how a wave function can be constructed that satisfies 
Eq.~(\ref{eq:3.1}) and for which the joint probability on the left-hand side 
of (\ref{eq:2.5})
is nonzero
\begin{equation}
P_{pp}(-,-) \geq 0 \ .
\label{eq:3.2}
\end{equation}
Let the wave function 
be denoted by $\psi_{\beta_1\beta_2}$, depending on the representation.
For example, $\psi_{xp}$ is the wave function in the representation
in which the position coordinate of particle 1 and the momentum
coordinate of particle 2 are employed.  Eq.~(\ref{eq:3.1}) imposes
the following conditions on the wave function:
\begin{eqnarray}
\psi_{xx}(x_1,x_2) & = & 0 
\ {\rm when} \ x_1 \geq 0 \ {\rm and} \ x_2 \geq 0 \ , \label{eq:3.3} \\   
\psi_{px}(p_1,x_2) & = & 0 
\ {\rm when} \ p_1 \leq 0 \ {\rm and} \ x_2 \leq 0 \ , \label{eq:3.4}
\end{eqnarray}
and
\begin{equation}    
\psi_{xp}(x_1,p_2) = 0 
\ {\rm when} \ x_1 \leq 0 \ {\rm and} \ p_2 \leq 0 \ . 
\label{eq:3.5}   
\end{equation}
A wave function satisfying these conditions can be constructed as follows:  
Let $g(p_1,p_2)$ be a function that is nonzero
only when $p_1$ and $p_2$ are positive, i.e.,
\begin{equation}
g(p_1,p_2) = 0 \ {\rm if} \ p_1 \leq 0 \ {\rm or} \ p_2 \leq 0 \ .
\label{eq:3.6}
\end{equation}
Its Fourier transform, denoted by $f(x_1,x_2)$, is
\begin{equation}
f(x_1,x_2) = \frac{1}{2\pi} \int_{-\infty}^\infty 
\int_{-\infty}^\infty e^{i(p_1x_1 + p_2x_2)} g(p_1,p_2) \ dp_1 dp_2 \ .
\label{eq:3.7}
\end{equation}
The wave function $\psi_{xx}$ is then given by
\begin{equation}
\psi_{xx}(x_1,x_2) = N[1 - \theta(x_1)\theta(x_2)] f(x_1,x_2)
\label{eq:3.8}
\end{equation}
where $\theta(x)$ is the Heaviside function defined by
\begin{equation}
\theta(x) = \left\{ \begin{array}{cc}
 1 & \ {\rm if} \ x \geq 0 \\
 0 & \ {\rm if} \ x < 0 \end{array} \right. 
\label{eq:3.9}
\end{equation}
and $N$ is the normalization coefficient chosen so that $\psi_{xx}(x_1,x_2)$
is normalized:
\begin{equation}
\int_{-\infty}^\infty \int_{-\infty}^\infty |\psi_{xx}(x_1,x_2)|^2 = 1 \ .
\label{eq:3.9b}
\end{equation}
Eq.~(\ref{eq:3.3}) is enforced by the factor in square brackets appearing
in Eq.~(\ref{eq:3.8}).  That Eq.~(\ref{eq:3.4}) is also satisfied is
now demonstrated.  $\psi_{px}$ is a Fourier transform of $\psi_{xx}$:
\begin{equation}
\psi_{px}(p_1,x_2) = \frac{1}{\sqrt{2\pi}} \int_{-\infty}^\infty
e^{-ip_1x_1} \psi_{xx}(x_1,x_2) \ dx_1 \ .
\label{eq:3.10}
\end{equation}
But, from Eq.~(\ref{eq:3.8}) this reduces to
\begin{equation}
\psi_{px}(p_1,x_2) = \frac{N}{\sqrt{2\pi}} \int_{-\infty}^\infty
e^{-ip_1x_1} f(x_1,x_2) \ dx_1
\label{eq:3.11}
\end{equation}
when $x_2 \leq 0$.
Substituting Eq.~(\ref{eq:3.7}) into this and carrying out the
$x_1$ integration followed by a momentum integration yields
\begin{equation}
\psi_{px}(p_1,x_2) = \frac{N}{\sqrt{2\pi}}
\int_{-\infty}^\infty e^{ip_2 x_2} g(p_1,p_2) dp_2
\label{eq:3.12}
\end{equation}
when $ x_2 \leq 0 $.  It is evident from Eq.~(\ref{eq:3.6}) that
the right-hand side of Eq.~(\ref{eq:3.12}) is zero when
$ p_1 \leq 0 $, that is, Eq.~(\ref{eq:3.4}) is satisfied.  
A similar argument shows that the wave function
of Eq.~(\ref{eq:3.8}) also satisfies Eq.~(\ref{eq:3.5}). 
Transforming Eq.~(\ref{eq:3.8}) into the momentum representation
for both particles yields, keeping Eq.~(\ref{eq:3.6}) in mind,
\begin{eqnarray}
\psi_{pp}(p_1,p_2) & = & - \frac{N}{2\pi} \int_0^\infty \int_0^\infty
e^{-i(p_1x_1 + p_2 x_2)}f(x_1,x_2) \ dx_1 dx_2 \nonumber \\
& & \ \ \ \ \  {\rm when} \ 
p_1 \leq 0 \ {\rm and} \ p_2 \leq 0 \ .
\label{eq:3.13}
\end{eqnarray}
$\psi_{pp}(p_1,p_2)$ evaluated over this range is what is needed to
compute $P_{pp}(-,-)$:
\begin{equation}
P_{pp}(-,-) = \int_{-\infty}^0 \int_{-\infty}^0
| \psi_{pp}(p_1,p_2)|^2 dp_1 dp_2 \ . 
\label{eq:3.14}
\end{equation}
If $\psi_{pp}(p_1,p_2) \neq 0$ over some region in the domain
($p_1 < 0 \ {\rm and} \ p_2 < 0$), then a wave function
has been constructed that violates the local realism condition 
Eq.~(\ref{eq:2.6}).

We now specialize to the case when $g(p_1,p_2)$ factorizes as
follows:
\begin{equation}
g(p_1,p_2) = g(p_1)g(p_2) 
\label{eq:4.1}
\end{equation}
where
\begin{equation}
g(p) = 0 \ {\rm for} \ p \leq 0 \ .
\label{eq:4.2}
\end{equation}
Then $f(x_1,x_2)$ factorizes,
\begin{equation}
f(x_1,x_2) = f(x_1)f(x_2) \ ,
\label{eq:4.3}
\end{equation}
where
\begin{equation}
f(x) = \frac{1}{\sqrt{2 \pi}} \int_0^\infty e^{ipx} g(p) dp \ .
\label{eq:4.4}
\end{equation}
Also, Eq.~(\ref{eq:3.13}) reduces to 
\begin{equation}
\psi_{pp}(p_1,p_2) = -N\psi_p(p_1)\psi_p(p_2) \ {\rm when} \ 
p_1 \leq 0 \ {\rm and} \ p_2 \leq 0 
\label{eq:4.5}
\end{equation}
where
\begin{equation}
\psi_p(p) = \frac{1}{\sqrt{2\pi}} \int_0^\infty e^{-ipx} f(x) dx \ 
{\rm when} \ p \leq 0 \ .
\label{eq:4.6}
\end{equation}
Substituting Eq.~(\ref{eq:4.5}) into Eq.~(\ref{eq:3.14}) yields
\begin{equation}
P_{pp}(-,-) = N^2\left[\int_{-\infty}^0 | \psi_p(p) |^2 dp \right]^2 \ .
\label{eq:4.7}
\end{equation}

As a specific example, let $g(p)$ be given by
\begin{equation}
g(p) = \left\{ \begin{array}{cc}
\sqrt{2\lambda} e^{-\lambda p} & \ {\rm for} \ p > 0 \\
0 & \ {\rm for } \ p \leq 0 
\end{array} \right. \ .
\label{eq:6.1}
\end{equation}
From this, using Eq.~(\ref{eq:4.4}), one obtains
\begin{equation}
f(x) = i \sqrt{\frac{\lambda}{\pi}} \frac{1}{x + i\lambda} \ .
\label{eq:6.2}
\end{equation}
Substituting this into Eq.~(\ref{eq:3.8}), using Eq.~(\ref{eq:4.3}), 
and computing the norm,
one obtains
\begin{equation}
N = \frac{2}{\sqrt{3}} \ .
\label{eq:6.3}
\end{equation}
Substituting Eq.~(\ref{eq:6.2}) into Eq.~(\ref{eq:4.6}) yields,
for $p \leq 0$,
\begin{equation}
\psi_p(p) = \frac{i}{\pi}\sqrt{\frac{\lambda}{2}}
\left[ \int_0^\infty \frac{\cos(|p|x)}{x + i\lambda} dx
+ i\int_0^\infty \frac{\sin(|p|x)}{x + i\lambda} dx \right] \ .
\label{eq:6.5}
\end{equation}
By breaking the right-hand side of this equation into real and
imaginary parts and by making use of formulas given by
Gradshteyn and Ryzhik \cite{gradshteyn} (section 3.723, Eqs. 1 through 4), 
this equation simplifies to
\begin{equation}
\psi_p(p) = - \frac{i}{\pi}\sqrt{\frac{\lambda}{2}}
e^{\lambda|p|}{\rm Ei}(-\lambda|p|) \ .
\label{eq:6.7}
\end{equation}
From this one finds
\begin{equation}
\int_{-\infty}^0 |\psi_p(p)|^2 dp = \frac{1}{2\pi^2}
\int_0^\infty e^{2x}{\rm Ei}^2(-x)dx \ .
\label{eq:6.8}
\end{equation}
By performing a numerical integration of this equation we have found that
\begin{equation}
\int_{-\infty}^0 |\psi_p(p)|^2 dp = \frac{1}{8}
\label{eq:6.13}
\end{equation}
to one part in $10^8$.
From Eqs.~(\ref{eq:4.7}) and (\ref{eq:6.3}) one thus obtains
\begin{equation}
P_{pp}(-,-) = \frac{1}{48} \ .
\label{eq:6.14}
\end{equation}
Thus, for a system possessing the wave function described here, a 
local realism violating event in which the momenta of both
particles are negative occurs at a rate of one event in 48 events.

It has been shown here that local realism violating experiments
of the Hardy type are, in principle, possible in which only position and
momentum measurements are performed.  A means of experimentally generating 
the appropriate states has
not been offered, so it remains to be seen whether such states can 
be realized in practice.  In this regard we derive hope from
the fact that the experiment proposed by EPR became realizable
57 years later as an optical analogue (through the development of parametric 
down-converters and homodyne detectors) and we take heart in the fact 
that state synthesis is an active topic of research \cite{vogel93}.

\end{document}